\documentclass[a4paper,11pt]{article}

\usepackage{amsmath, amsfonts, url}
\usepackage{graphicx}
\graphicspath{{fig/}}
\usepackage{a4wide}
\usepackage[parfill]{parskip}
\usepackage{geometry}
\usepackage{pgf,tikz}
\usetikzlibrary{arrows}
\usepackage{rotating}
\usepackage{verbatim}
\usepackage[parfill]{parskip}
\usepackage{natbib}
\usepackage[ruled]{algorithm2e}

\newtheorem{definition}{Definition}

\newcommand{\erdos}{GER }
\newcommand{\hetero}{two-colour \erdos}
\newcommand{\ie}{\emph{i.e.}, }

\newcommand{\eg}{\emph{e.g.}, }

\title{All networks look the same to me: \\ Testing for homogeneity in networks}

\author{Jono Tuke and Matthew Roughan}

\date{\today}

\begin{document}
	\maketitle

	\begin{abstract}

		How can researchers test for heterogeneity in the local structure of a
		network? In this paper, we present a framework that utilizes random sampling
		to give subgraphs which are then used in a goodness of fit test to test for
		heterogeneity. We illustrate how to use the goodness of fit test for an
		analytically derived distribution as well as an empirical distribution. To
		demonstrate our framework, we consider the simple case of testing for edge
		probability heterogeneity. We examine the significance level, power and
		computation time for this case with appropriate examples. Finally we outline
		how to apply our framework to other heterogeneity problems.

	\end{abstract}

\section{Introduction}
\label{sec:Introduction}

There are many examples of complex-interaction systems, often
described as {\em networks}, for which we can have only a single
example: \eg
\begin{itemize}
\item the phylogenetic tree describing the evolution of species \citep{huelsenbeck01:_mrbay};
\item the Internet \citep{roughan11};
\item the global inter-species food web \citep{Dunne01102002};
\item the scientific collaboration network \citep{Newman16012001}; or
\item the complete human social network \citep{wasserman94:_social_networ_analy}.
\end{itemize}
Although these networks are sometimes considered as pluralities (as the
references above often do), there is actually a single network from
which we observe smaller components. The network might evolve over
time, but snapshots of its evolution are highly dependent samples of
the network in question. So in reality, we have one sample of each.

There is a philosophical problem in modelling a system for which we
have a single data point, let alone a high-dimensional system with
only one datum. Namely, how can we balance the conflicting demands in
modelling: on the one hand we want a model that correctly fits the
observations; and on the other hand a model that is simple and
explanatory. The former extreme is represented by ``the data is the
model'' (which holds zero explanatory power), and the later by the
model that all datasets are the same, and any discrepancy is just
noise (a ludicrous supposition, presented only as a example of the
extreme).

Somewhere in between lies what Box and Draper would call a ``useful''
model \citep{box87:_models}, but how to know?  Statistics has been
employed for more than 200 years towards answering that very
question. For instance, a model might commonly be tested through some
form of cross-validation where some data-points are removed from the
fitting process in order to provide a test set. If we have only one
datum this is clearly impossible, as are all simple statistical tests.

Consequently, the analysis and modelling of such systems can take one
of two forms\footnote{Of course there are other approaches, but they
  often involve flawed logic, \ie model X has feature A, and we
  observe A, therefore model X is correct.}:

\begin{enumerate}
\item We know the ``physics'' of the system, at least to some
  approximation, and exploit this side-information in our
  analysis. For instance, knowing that a network was generated by a
  process of growth with preferential
  attachment~\citep{barabasi99:_scale_free}, we could use the data
  simply to estimate the growth parameters\footnote{Note that using
    the data for parameter estimation from a model is very different
    from using the data to choose the model!}.

\item Or we make the critical assumption that smaller components of
  the model can be used to obtain samples from which we can draw
  statistics, and thereby make an assessment of the correct model. For
  instance, in the analysis of food webs, we might examine
  (approximately) isolated groups, \eg \citet{Dunne01102002}.
\end{enumerate}

Of the two methods, the second approach is valued for its explanatory
power, \ie from it we can derive new physics.  However, the critical
assumption must be valid.  In the analysis of subpopulations of
species the assumption may well be, as it might in a network that can
decompose into only loosely coupled components, each of which can be
used to provide pseudo-independent samples of the larger network. But
sometimes, the assumption is worrying.

The assumption requires that local structure of a network is
homogenous. In our context that means that each subgraph is
statistically the same. If not, we have the possibility of what is
sometimes called a Type-III error ``giving the right answer to the
wrong problem'' \citep{Kimball:1957ba}. That is, a perfectly
reasonable statistical procedure may provide results that would be
correct given the assumption, but are actually far from reasonable.

So how can we test the relationship between the global structure of a
network and the local structure of its subgraphs? Let's illustrate
with a simple example of the Gilbert-Erd\"os-R\'enyi (GER) network. In
\erdos networks, the probability of an edge between any two nodes is
constant (See Section~\ref{sec:Sampling}).  What if we have a network
where the probability of an edge depends on a property of the
nodes. Now the subgraphs will have heterogeneity in the overall
probability of edges.

How do we test for this heterogeneity, how do examine the relationship
between the local structure of subgraphs and the global structure of a
network?

In this paper, we illustrate how this is achieved using sampling from
the network. This paper is a proof of concept with a simple example to
illustrate its potential. We consider two simple tests to test for
homogeneity of edge probability. These methods use sampling without
replacement and goodness of fit tests. We outline the algorithm for
both methods and test their significance, power, and computational
cost and show that they can provide practical analysis of random
networks.

Ultimately, the goal is to provide a path towards diagnostic tests of
network models. That is, rather than approaching modelling as a craft,
requiring expertise and deep knowledge of graph theory, we should be
able to provide a standard suite of estimators, along with diagnostics
for those estimators that can be used by any practitioner, much as
modern statistics has for regression. For instance, on performing
linear regression, one might then test for heteroscedasity. Here we
propose fitting random-graph models, but with formal means to test
underlying assumptions inherent in the test.

The exemplars we present are simple, but the advantage of the
underlying idea (as opposed to a test designed specifically for a
particular model) is that it is easily generalised; it is simply a
matter of choosing an appropriate sampling, and variate against which
to measure homogeneity. Then one can consider and test for more
general notions of heterogeneity.

\section{Graphs and Sampling}
\label{sec:Sampling}

Consider an undirected network $(V,E)$ such that the $(i,j)$ entry of
the adjacency matrix is denoted $A_{ij}$ and is defined as
\[
A_{ij} = \begin{cases}
  1, &\text{ edge between node $i$ and node $j$},\\
  0, &\text{ no edge between node $i$ and node $j$}.
\end{cases}
\]
The networks we consider are undirected with no loops so $A_{ij} = A_{ji}$, and
$A_{ii} = 0$, but the results are generalizable. Denote the number of
edges and nodes as $|E|$ and $|V|$, respectively.

We denote by $p_{ij}$ the probability $P\{A_{ij} = 1\}$, and make the
following definition:

\begin{definition}
  We call a network {\em homogeneous with respect to edge probability}
  if the edge probability $p_{ij} = p$ is constant, \ie the edge
  probability does not depend on any (potentially hidden) node
  properties.
\end{definition}

Note two features of the definition:
\begin{enumerate}

\item We define homogeneity with respect to a feature of the graph. A
  common source of confusion in the term seems to be that different
  authors use homogeneity with respect to alternative features: here
  we suggest that the definition must be explicit, but the definition
  generalizes in the sense that we could easily incorporate other
  variates in place of edge probability.

\item The definition here is equivalent to that of \erdos random
  graphs $G(n,p)$, but that is not the aim. Our goal is to test for a
  {\em feature} of the data, not a specific model.

\end{enumerate}
The latter is an important point in general. Much statistical
modelling is about fitting models or estimating parameters, given
certain assumptions. This paper is aimed at testing assumptions.  The
distinction is important: as a result we will not present the common
approach to test for a \erdos graph by examining the node degree
distribution, as calculation of this distribution inherently presumes
{\em a priori} that the network is homogenous, and that we can
estimate the distribution by examining the statistics of all nodes as
if they were identically distributed samples drawn from an underlying
variate.




We present two methods for testing if an observed network is
homogeneous with respect to edge probability, \ie
\begin{align*}
	H_0 &: p_{ij} = p\\
	H_a &: \text{at least one } p_{ij}\neq p.
\end{align*}
Note, however, that we only know the $A_{ij}$, not the $p_{ij}$. We can't
even form reasonable estimates of the individual $p_{ij}$, as we have
a single sample of each.


Both methods we present use a goodness of fit test applied to the
number of edges observed in sampled subgraphs from the observed
network. To obtain a sampled subgraph, $G'(V', E')$ we sample $k = |V
'|$ nodes (without replacement) from the observed network $G(V,E)$ and
choose $E'$ such that
\[
(i,j) \in E', \text{iff } i,j \in V' \text{ and }(i,j)\in E.
\]

The strategy above is referred to as {\em node sampling} (see
\cite{PhysRevE.73.016102}).  However, once again, note that the
sampling procedure is arbitrary. We have chosen one of the simplest
possible here, but there are alternative approaches such as link and
snowball sampling \cite{PhysRevE.73.016102}.

\section{Testing for Homogeneity}
\label{sec:Comparing samples to expectation}

The statistics of sampled subgraphs have been studied, however, that
study seems to have focussed on standard graph metrics, such as
power-law degree exponents, \eg see \cite{PhysRevE.73.016102}. Here
we are interested in statistics that are actually somewhat simpler,
but more general (than a metric that presumes a certain model).

We denote the number of edges in the $n$th sampled subgraph by the random
variable $Y_n$. For the goodness of fit test, we require its distribution. In
the observed network we have $|V|$ nodes which gives $V_e = {|V| \choose 2}$
possible edges, of which $|E|$ exist. If we have a sampled subgraph of $k$ nodes
this has $k_e = {k \choose 2}$ possible edges. Naively we might estimate the
probability of observing $y_n$: edge in the $n$th subgraph to be
\[
    P(Y_n = y_n) = \frac{{|E|\choose y}{V_e - |E| \choose k_e - y_n}}{{V_e\choose k_e}}.
\]

This probability comes from the fact that the number of ways of choosing the
$y_n$ edges from the total $|E|$ edges in the observed network is ${|E| \choose
2}$; we must also have chosen $k_e - y_n$ potential edges that were not edges -
there are ${V_e - |E| \choose k_e - y_n}$ ways of doing this. Therefore
${|E|\choose y}{V_e - |E| \choose k_e - y_n}$ is the number of ways we can take
a sample of $k$ nodes with $y_n$ edges. The total number of samples of size $k$
nodes that we can take from the observed network is ${V_e\choose k_e}$, and
hence the probability above. Observant readers will notice that this is the
probability mass function of a hypergeometric distribution with a population of
size $V_e$ with $|E|$ successes and $V_e - |E|$ failures, from which we take a
random sample of $k_e$ without replacement and denote the number of successes in
the sample as $Y_n$.

However, the analysis above assumes that the edges are independent of one
another in the sampling process, \ie, the selection of an edge does not change
the probability of another edge being selected in the same random sample. This
is not the case as we will now illustrate.

\begin{figure}[htbp]
	\centering
	\begin{tikzpicture}[node distance = 3cm]

		\tikzstyle{state}=[draw, black, circle, minimum size=30pt, inner sep=5pt]

		\node[state] (a1) {$1$};
		\node[state] (a2) [right of=a1] {$2$};
		\node[state] (a3) [below of=a1] {$3$};
		\node[state] (a4) [below of=a2] {$4$};

		\node[state] (b1) [right of=a2] {$1$};
		\node[state] (b2) [right of=b1] {$2$};
		\node[state] (b3) [below of=b1] {$3$};
		\node[state] (b4) [below of=b2] {$4$};

		\node (a) at (1.5,-4) {(a)};
		\node (a) at (7.5,-4) {(b)};

		\path[-, very thick]
		(a1) edge (a2)
		edge (a3);

		\path[-, very thick]
		(b1) edge (b2);

		\path[-, very thick]
		(b3) edge (b4);

	\end{tikzpicture}
	\caption{Two networks illustrating that sampling from networks does not give a hypergeometric distribution.}
	\label{fig:small_graph}
\end{figure}
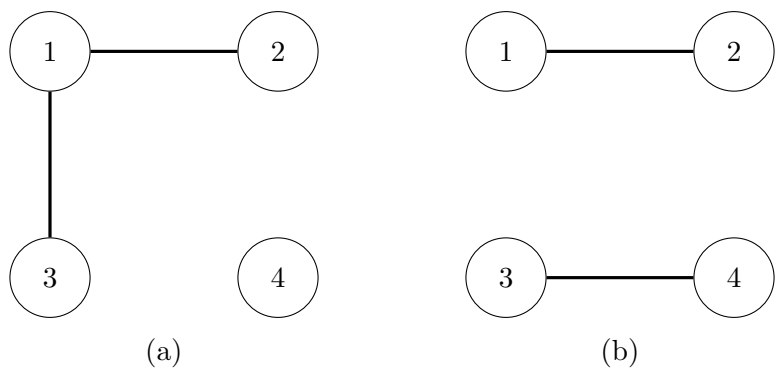

Consider the two networks given in Figure~\ref{fig:small_graph} denoted Network~(a) and Network~(b). Both of these networks have 4 nodes and two edges. Consider sampling three nodes from each network. The possible samples are given in Table~\ref{tab:small_graph_edges}.

\begin{table}
	\begin{center}
		\begin{tabular}{c|r|r}
			& \multicolumn{2}{c}{Number of edges }\\
			Nodes sampled & Network (a) & Network (b)\\
			\hline
			$\{1,2,3\}$ & 2 & 1\\
			$\{1,2,4\}$ & 1 & 1\\
			$\{1,3,4\}$ & 1 & 1\\
			$\{2,3,4\}$ & 0 & 1\\
		\end{tabular}
		\caption{Edges obtained for each of the possible samples of three nodes from Networks (a) and (b) in Figure~\ref{fig:small_graph}.}
		\label{tab:small_graph_edges}
	\end{center}
\end{table}

\begin{table}
	\begin{center}
		\begin{tabular}{r|rrr}
			& \multicolumn{3}{c}{$y$}\\ 
			           & 0 & 1 & 2 \\\hline
			Network (a)& 1/4 & 1/2 & 1/4 \\
			Network (b)& 0 & 1 & 0 \\
			Hypergeometric(2,4,3)& 1/5 & 3/5 & 1/5 \\
		\end{tabular}
		\caption{The probability mass function $P(Y=y)$ for
                  the number of edges in a sampled subgraph of size 3
                  from the networks in
                  Figure~\ref{fig:small_graph}. For reference the
                  theoretical probabilities calculated from the
                  equivalent hypergeometric distribution are given.}
		\label{tab:small_graph_pmf}
	\end{center}
\end{table}

\begin{figure}[htbp]
	\centering
	\begin{tikzpicture}[node distance = 1.5cm]

		\tikzstyle{state}=[draw, black, circle]

		\node[state] (a1) {$1$};
		\node[state] (a2) [right of=a1] {$2$};
		\node[state] (a3) [below of=a1] {$3$};
		\node[state] (a4) [below of=a2] {$4$};

		\path[-, very thick]
		(a1) edge (a2)
		(a1) edge (a3);

		\node[state] (b1) [right of=a2] {$1$};
		\node[state] (b2) [right of=b1] {$2$};
		\node[state] (b3) [below of=b1] {$3$};
		\node[state] (b4) [below of=b2] {$4$};

		\path[-,very thick]
		(b1) edge (b2)
		(b1) edge (b4);

		\node[state] (c1) [right of=b2] {$1$};
		\node[state] (c2) [right of=c1] {$2$};
		\node[state] (c3) [below of=c1] {$3$};
		\node[state] (c4) [below of=c2] {$4$};

		\path[-, very thick]
		(c1) edge (c2)
		(c3) edge (c2);

		\node[state] (d1) [right of=c2] {$1$};
		\node[state] (d2) [right of=d1] {$2$};
		\node[state] (d3) [below of=d1] {$3$};
		\node[state] (d4) [below of=d2] {$4$};

		\path[-, very thick]
		(d1) edge (d2)
		(d2) edge (d4);

		\node[state] (e1) [right of=d2] {$1$};
		\node[state] (e2) [right of=e1] {$2$};
		\node[state] (e3) [below of=e1] {$3$};
		\node[state] (e4) [below of=e2] {$4$};

		\path[-, very thick]
		(e1) edge (e2)
		(e3) edge (e4);

		\node[state] (f1) [below of=a3] {$1$};
		\node[state] (f2) [right of=f1] {$2$};
		\node[state] (f3) [below of=f1] {$3$};
		\node[state] (f4) [below of=f2] {$4$};

		\path[-, very thick]
		(f1) edge (f3)
		(f1) edge (f4);

		\node[state] (g1) [right of=f2] {$1$};
		\node[state] (g2) [right of=g1] {$2$};
		\node[state] (g3) [below of=g1] {$3$};
		\node[state] (g4) [below of=g2] {$4$};

		\path[-, very thick]
		(g1) edge (g3)
		(g3) edge (g2);

		\node[state] (h1) [right of=g2] {$1$};
		\node[state] (h2) [right of=h1] {$2$};
		\node[state] (h3) [below of=h1] {$3$};
		\node[state] (h4) [below of=h2] {$4$};

		\path[-, very thick]
		(h1) edge (h3)
		(h4) edge (h2);

		\node[state] (i1) [right of=h2] {$1$};
		\node[state] (i2) [right of=i1] {$2$};
		\node[state] (i3) [below of=i1] {$3$};
		\node[state] (i4) [below of=i2] {$4$};

		\path[-, very thick]
		(i1) edge (i3)
		(i4) edge (i3);

		\node[state] (j1) [right of=i2] {$1$};
		\node[state] (j2) [right of=j1] {$2$};
		\node[state] (j3) [below of=j1] {$3$};
		\node[state] (j4) [below of=j2] {$4$};

		\path[-, very thick]
		(j1) edge (j4)
		(j2) edge (j3);

		\node[state] (k1) [below of=f3] {$1$};
		\node[state] (k2) [right of=k1] {$2$};
		\node[state] (k3) [below of=k1] {$3$};
		\node[state] (k4) [below of=k2] {$4$};

		\path[-, very thick]
		(k1) edge (k4)
		(k2) edge (k4);

		\node[state] (l1) [right of=k2] {$1$};
		\node[state] (l2) [right of=l1] {$2$};
		\node[state] (l3) [below of=l1] {$3$};
		\node[state] (l4) [below of=l2] {$4$};

		\path[-, very thick]
		(l1) edge (l4)
		(l3) edge (l4);

		\node[state] (m1) [right of=l2] {$1$};
		\node[state] (m2) [right of=m1] {$2$};
		\node[state] (m3) [below of=m1] {$3$};
		\node[state] (m4) [below of=m2] {$4$};

		\path[-, very thick]
		(m2) edge (m3)
		(m2) edge (m4);

		\node[state] (n1) [right of=m2] {$1$};
		\node[state] (n2) [right of=n1] {$2$};
		\node[state] (n3) [below of=n1] {$3$};
		\node[state] (n4) [below of=n2] {$4$};

		\path[-, very thick]
		(n2) edge (n3)
		(n3) edge (n4);
		\path[-, very thick]
		(n2) edge (n3)
		(n3) edge (n4);

		\node[state] (o1) [right of=n2] {$1$};
		\node[state] (o2) [right of=o1] {$2$};
		\node[state] (o3) [below of=o1] {$3$};
		\node[state] (o4) [below of=o2] {$4$};

		\path[-, very thick]
		(o2) edge (o4)
		(o3) edge (o4);
	\end{tikzpicture}
	\caption{All possible networks of four nodes with two
          edges. They are all isomorphic to the networks in
          Figure~\ref{fig:small_graph}: twelve to Network (a), and
          three to Network (b).}
	\label{fig:small_graph_all}
\end{figure}
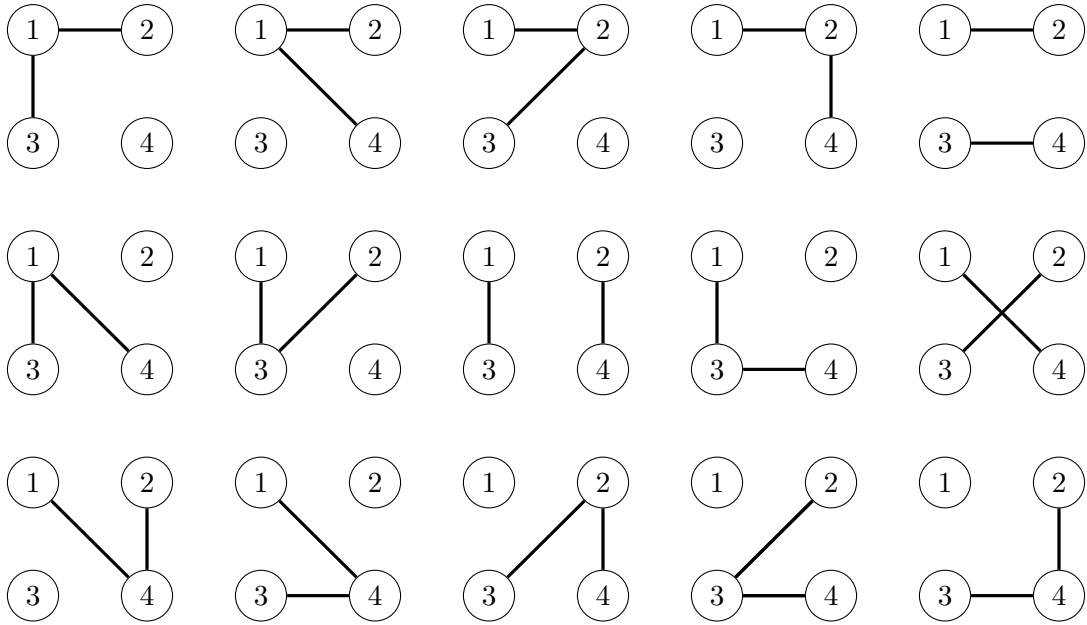

The resultant probability mass functions and reference hypergeometric are given
in Table~\ref{tab:small_graph_pmf}. The distributions are not equivalent to the
hypergeometric and not equivalent to one another. So where is the hypergeometric
distribution? Consider all possible networks of four nodes with two edges given
in Figure~\ref{fig:small_graph_all}. These fifteen networks are all isomorphic
to Network~(a) or Network~(b). Assuming that we know we have a network of four
nodes and two edges and sample three nodes, the probabilities  $P(Y = y
\mid \text{Network }X)$ are given in
Table~\ref{tab:small_graph_pmf}, and we can then use the Law of Total
Probability to show that $P(Y = y)$ is the same as that given for the
hypergeometric distribution given in Table~\ref{tab:small_graph_edges}. Thus the discrepancy arises because we sample from a single \emph{real} graph, not the complete ensemble of possible random graphs.

As will be seen later, this discrepancy between the true distribution of the
number of edges in a subgraph and the hypergeometric distribution becomes
negligible as the number of nodes in the observed network becomes large. This
means that for large networks, we can use the equivalent hypergeometric
distribution to calculated the expected number of edges for the goodness of fit
test for homogeneity. The process is outlined in Algorithm~\ref{algo:gof_large}.
We refer to this method as the \emph{Approximation test}.

\begin{algorithm}[htbp]
	\LinesNumbered
	\SetKwInOut{Input}{Input}
	\Input{Observed network $G(V,E)$}
	\BlankLine

        \For{$n = 1, \dots, N$}{
          Sample a subgraph with $k$ nodes from the network $G$.

          Count the number of edges: $y_n$

        }

	Tabulate $y_n$ into bins according to Algorithm~\ref{algo:bin} to give the number of subgraphs in each bin: $f_m, m = 1, \dots, M$.

	Calculate the expected number in each bin from the equivalent hypergeometric: $e_m, m = 1, \dots, M$.

	Calculate the $\chi^2$ statistic:
	\[
	X^2 = \sum^M_{m = 1} \frac{(f_m - e_m)^2}{e_m}.
	\]\label{step:chi}

	Calculate the P-value:
	\[
	P(X^2 \geq x^2), \text{ where } X^2\sim\chi^2_{M-1}.
	\]\label{step:pv}

	\Return{P-value}

	\caption{Goodness of fit test for homogeneity in large networks}\label{algo:gof_large}
\end{algorithm}

For small networks, we modify Algorithm~\ref{algo:gof_large} so that instead of
calculating the P-value using a $\chi^2$ distribution as given in
Step~\ref{step:pv}, we simulate an empirical distribution of $X^2$. This is
achieved by simulating $R$ \erdos networks with the same number of nodes and
edges as the observed network. For each of these simulated networks, we use
Algorithm~\ref{algo:gof_large} to calculate $x^2_r, r = 1, \dots, R$, where
$x^2_r$ is the observed value of $X^2$ (Step~\ref{step:chi} of
Algo.~\ref{algo:gof_large}) of the $r$th network. The empirical P-value,
P-value$_{emp}$ is then defined

\[
\text{P-value}_{emp} = \frac1R\sum^R_{r = 1} I(x^2_r \geq x^2_{obs}),
\]

where $x^2_{obs}$ is the observed value of $X^2$ for the observed network, and
$I(\cdot)$ is an indicator function. This modification we will refer to as the
\emph{Empirical Test}. Its advantage is that it is more accurate for small samples, but as we will show it is much less computationally efficient.

\section{Results}
\label{sec:Results}

\subsection{Significance levels}
\label{sec:sign}

We examined the ability of the proposed methods -- the Approximation
Test and the Empirical Test -- to identify networks with
hetereogeneity by assessing the significance level and power for both
methods. To test the significance level, we simulated \erdos networks
of size $|V| = 10^i, i = \{2.00, 2.25, \ldots, 4.00\}$ with average
node degree, denoted $\bar{d}$, of 1, 3, 5, and 10 (sparse graphs are
more realistic for many applications, but also sparsity makes the
estimation problem harder, so we test in this domain). For each of
these pairs of parameters $(|V|,\bar{d})$ we simulate 500 \erdos
networks, and for each of these networks we sampled 1000 subgraphs of
size $k$.

The size $k$ was chosen to maximise the variance of the number of
edges for the $i$th subgraph, $Y_n$, under the assumption that $Y_n$
has a hypergeometric distribution. It can be shown that this is
achieved by
\[
k = \frac{1 + \sqrt{1 + 2 \times |V|(|V| - 1)}}{2}.
\]

We assessed the Empirical Test using the same parameter settings, with
the exception that we considered small networks ($|V| \leq 10^3$)
only, due to the much larger computation times for this method (see
Section~\ref{sec:Computational time} for details). We also reduced the
number of replications from 500 to 200 as a result of the larger
computation times. We use 200 samples to simulate the empirical
distribution of $X^2$ in this test: this proved to be a reasonable
compromise between accuracy and computation time.

\subsection{Significance} \label{sub:Significance}

For each parameter pair $(|V|, \bar{d})$ we calculated an estimated
significance level by calculating the proportion of the P-values and
P-value$_{emp}$ that were less than or equal to 0.05. These are shown
in Figure~\ref{fig:sign_level} for $\bar{d} = 5$ with 95\% confidence
intervals for the true significance level. The estimated significance
levels for the Empirical Test are not significantly different to 0.05,
as desired. On the other hand, the significance level for the
Approximation Test were significantly higher than expected for small
networks. For $|V| < 1000$ we see values as large as $0.25$,
indicating a false positive rate of 25\%. The significance levels drop
back to the correct range at around $|V| = 1000$.

Similar results were obtained for the other values of $\bar{d}$, as
can be see in Table~\ref{tab:sig_pv}. The net result is that one
should apply the Approximation Test only for larger networks ($|V|
\geq 1000$), but in that range the approximation works quite well.

\begin{figure}
	\centering
	\scalebox{0.9}
	{
	\input{./significance_level.tex}
	}
        \vspace{-6mm}
	\caption{Estimated significance level with 95\% confidence
          intervals for the true significance level for $\bar{d} =
          5$. Note that the Empirical Test's significance is within
          bounds of the desire level for all $|V|$, but the
          Approximation Test only approaches the correct significance
          near $|V| = 600$.}\label{fig:sign_level}
	\scalebox{0.9}
	{
\begin{tikzpicture}[x=1pt,y=1pt]
\definecolor{fillColor}{RGB}{255,255,255}
\path[use as bounding box,fill=fillColor,fill opacity=0.00] (0,0) rectangle (433.62,289.08);
\begin{scope}
\path[clip] (  0.00,  0.00) rectangle (433.62,289.08);
\definecolor{drawColor}{RGB}{255,255,255}
\definecolor{fillColor}{RGB}{255,255,255}

\path[draw=drawColor,line width= 0.6pt,line join=round,line cap=round,fill=fillColor] (  0.00,  0.00) rectangle (433.62,289.08);
\end{scope}
\begin{scope}
\path[clip] ( 40.27, 30.69) rectangle (428.12,283.58);
\definecolor{fillColor}{RGB}{255,255,255}

\path[fill=fillColor] ( 40.27, 30.69) rectangle (428.12,283.58);
\definecolor{drawColor}{gray}{0.98}

\path[draw=drawColor,line width= 0.6pt,line join=round] ( 40.27, 37.66) --
	(428.12, 37.66);

\path[draw=drawColor,line width= 0.6pt,line join=round] ( 40.27,109.46) --
	(428.12,109.46);

\path[draw=drawColor,line width= 0.6pt,line join=round] ( 40.27,181.25) --
	(428.12,181.25);

\path[draw=drawColor,line width= 0.6pt,line join=round] ( 40.27,253.05) --
	(428.12,253.05);

\path[draw=drawColor,line width= 0.6pt,line join=round] (146.04, 30.69) --
	(146.04,283.58);

\path[draw=drawColor,line width= 0.6pt,line join=round] (322.34, 30.69) --
	(322.34,283.58);
\definecolor{drawColor}{gray}{0.90}

\path[draw=drawColor,line width= 0.2pt,line join=round] ( 40.27, 73.56) --
	(428.12, 73.56);

\path[draw=drawColor,line width= 0.2pt,line join=round] ( 40.27,145.36) --
	(428.12,145.36);

\path[draw=drawColor,line width= 0.2pt,line join=round] ( 40.27,217.15) --
	(428.12,217.15);

\path[draw=drawColor,line width= 0.2pt,line join=round] ( 57.90, 30.69) --
	( 57.90,283.58);

\path[draw=drawColor,line width= 0.2pt,line join=round] (234.19, 30.69) --
	(234.19,283.58);

\path[draw=drawColor,line width= 0.2pt,line join=round] (410.49, 30.69) --
	(410.49,283.58);
\definecolor{drawColor}{RGB}{0,0,0}
\definecolor{fillColor}{RGB}{0,0,0}

\path[draw=drawColor,line width= 0.4pt,line join=round,line cap=round,fill=fillColor] ( 57.90, 42.18) circle (  1.96);

\path[draw=drawColor,line width= 0.4pt,line join=round,line cap=round,fill=fillColor] (102.04, 43.10) circle (  1.96);

\path[draw=drawColor,line width= 0.4pt,line join=round,line cap=round,fill=fillColor] (145.99, 45.78) circle (  1.96);

\path[draw=drawColor,line width= 0.4pt,line join=round,line cap=round,fill=fillColor] (190.07, 48.21) circle (  1.96);

\path[draw=drawColor,line width= 0.4pt,line join=round,line cap=round,fill=fillColor] (234.19, 54.78) circle (  1.96);

\path[draw=drawColor,line width= 0.4pt,line join=round,line cap=round,fill=fillColor] (278.26, 61.96) circle (  1.96);

\path[draw=drawColor,line width= 0.4pt,line join=round,line cap=round,fill=fillColor] (322.34, 72.73) circle (  1.96);

\path[draw=drawColor,line width= 0.4pt,line join=round,line cap=round,fill=fillColor] (366.41, 86.97) circle (  1.96);

\path[draw=drawColor,line width= 0.4pt,line join=round,line cap=round,fill=fillColor] (410.49,104.69) circle (  1.96);

\path[draw=drawColor,line width= 0.4pt,line join=round,line cap=round,fill=fillColor] ( 57.90,207.71) circle (  1.96);

\path[draw=drawColor,line width= 0.4pt,line join=round,line cap=round,fill=fillColor] (102.04,208.93) circle (  1.96);

\path[draw=drawColor,line width= 0.4pt,line join=round,line cap=round,fill=fillColor] (145.99,211.06) circle (  1.96);

\path[draw=drawColor,line width= 0.4pt,line join=round,line cap=round,fill=fillColor] (190.07,214.37) circle (  1.96);

\path[draw=drawColor,line width= 0.4pt,line join=round,line cap=round,fill=fillColor] (234.19,219.91) circle (  1.96);

\path[draw=drawColor,line width= 0.4pt,line join=round,line cap=round,fill=fillColor] (278.26,228.36) circle (  1.96);

\path[draw=drawColor,line width= 0.4pt,line join=round,line cap=round,fill=fillColor] (322.34,239.54) circle (  1.96);

\path[draw=drawColor,line width= 0.4pt,line join=round,line cap=round,fill=fillColor] (366.41,253.92) circle (  1.96);

\path[draw=drawColor,line width= 0.4pt,line join=round,line cap=round,fill=fillColor] (410.49,272.08) circle (  1.96);

\path[draw=drawColor,line width= 0.6pt,line join=round] ( 57.90, 42.18) --
	(102.04, 43.10) --
	(145.99, 45.78) --
	(190.07, 48.21) --
	(234.19, 54.78) --
	(278.26, 61.96) --
	(322.34, 72.73) --
	(366.41, 86.97) --
	(410.49,104.69);

\path[draw=drawColor,line width= 0.6pt,dash pattern=on 2pt off 2pt ,line join=round] ( 57.90,207.71) --
	(102.04,208.93) --
	(145.99,211.06) --
	(190.07,214.37) --
	(234.19,219.91) --
	(278.26,228.36) --
	(322.34,239.54) --
	(366.41,253.92) --
	(410.49,272.08);
\definecolor{drawColor}{gray}{0.50}

\path[draw=drawColor,line width= 0.6pt,line join=round,line cap=round] ( 40.27, 30.69) rectangle (428.12,283.58);
\end{scope}
\begin{scope}
\path[clip] (  0.00,  0.00) rectangle (433.62,289.08);
\definecolor{drawColor}{RGB}{0,0,0}

\node[text=drawColor,anchor=base east,inner sep=0pt, outer sep=0pt, scale=  0.88] at ( 35.32, 70.53) {0.1};

\node[text=drawColor,anchor=base east,inner sep=0pt, outer sep=0pt, scale=  0.88] at ( 35.32,142.33) {1.0};

\node[text=drawColor,anchor=base east,inner sep=0pt, outer sep=0pt, scale=  0.88] at ( 35.32,214.12) {10.0};
\end{scope}
\begin{scope}
\path[clip] (  0.00,  0.00) rectangle (433.62,289.08);
\definecolor{drawColor}{RGB}{0,0,0}

\path[draw=drawColor,line width= 0.6pt,line join=round] ( 37.52, 73.56) --
	( 40.27, 73.56);

\path[draw=drawColor,line width= 0.6pt,line join=round] ( 37.52,145.36) --
	( 40.27,145.36);

\path[draw=drawColor,line width= 0.6pt,line join=round] ( 37.52,217.15) --
	( 40.27,217.15);
\end{scope}
\begin{scope}
\path[clip] (  0.00,  0.00) rectangle (433.62,289.08);
\definecolor{drawColor}{RGB}{0,0,0}

\path[draw=drawColor,line width= 0.6pt,line join=round] ( 57.90, 27.94) --
	( 57.90, 30.69);

\path[draw=drawColor,line width= 0.6pt,line join=round] (234.19, 27.94) --
	(234.19, 30.69);

\path[draw=drawColor,line width= 0.6pt,line join=round] (410.49, 27.94) --
	(410.49, 30.69);
\end{scope}
\begin{scope}
\path[clip] (  0.00,  0.00) rectangle (433.62,289.08);
\definecolor{drawColor}{RGB}{0,0,0}

\node[text=drawColor,anchor=base,inner sep=0pt, outer sep=0pt, scale=  0.88] at ( 57.90, 19.68) {100};

\node[text=drawColor,anchor=base,inner sep=0pt, outer sep=0pt, scale=  0.88] at (234.19, 19.68) {1000};

\node[text=drawColor,anchor=base,inner sep=0pt, outer sep=0pt, scale=  0.88] at (410.49, 19.68) {10000};
\end{scope}
\begin{scope}
\path[clip] (  0.00,  0.00) rectangle (433.62,289.08);
\definecolor{drawColor}{RGB}{0,0,0}

\node[text=drawColor,anchor=base,inner sep=0pt, outer sep=0pt, scale=  1.10] at (234.19,  7.70) {$|V|$};
\end{scope}
\begin{scope}
\path[clip] (  0.00,  0.00) rectangle (433.62,289.08);
\definecolor{drawColor}{RGB}{0,0,0}

\node[text=drawColor,rotate= 90.00,anchor=base,inner sep=0pt, outer sep=0pt, scale=  1.10] at ( 15.28,157.13) {Computation time (sec)};
\end{scope}
\begin{scope}
\path[clip] (  0.00,  0.00) rectangle (433.62,289.08);
\definecolor{fillColor}{RGB}{255,255,255}

\path[fill=fillColor] ( 70.70,233.97) rectangle (164.98,282.61);
\end{scope}
\begin{scope}
\path[clip] (  0.00,  0.00) rectangle (433.62,289.08);
\definecolor{drawColor}{RGB}{0,0,0}

\node[text=drawColor,anchor=base west,inner sep=0pt, outer sep=0pt, scale=  1.10] at ( 74.97,270.76) {Method};
\end{scope}
\begin{scope}
\path[clip] (  0.00,  0.00) rectangle (433.62,289.08);
\definecolor{drawColor}{gray}{0.80}
\definecolor{fillColor}{RGB}{255,255,255}

\path[draw=drawColor,line width= 0.6pt,line join=round,line cap=round,fill=fillColor] ( 74.97,252.70) rectangle ( 89.42,267.15);
\end{scope}
\begin{scope}
\path[clip] (  0.00,  0.00) rectangle (433.62,289.08);
\definecolor{drawColor}{RGB}{0,0,0}

\path[draw=drawColor,line width= 0.6pt,line join=round] ( 76.41,259.92) -- ( 87.98,259.92);
\end{scope}
\begin{scope}
\path[clip] (  0.00,  0.00) rectangle (433.62,289.08);
\definecolor{drawColor}{gray}{0.80}
\definecolor{fillColor}{RGB}{255,255,255}

\path[draw=drawColor,line width= 0.6pt,line join=round,line cap=round,fill=fillColor] ( 74.97,238.24) rectangle ( 89.42,252.70);
\end{scope}
\begin{scope}
\path[clip] (  0.00,  0.00) rectangle (433.62,289.08);
\definecolor{drawColor}{RGB}{0,0,0}

\path[draw=drawColor,line width= 0.6pt,dash pattern=on 2pt off 2pt ,line join=round] ( 76.41,245.47) -- ( 87.98,245.47);
\end{scope}
\begin{scope}
\path[clip] (  0.00,  0.00) rectangle (433.62,289.08);
\definecolor{drawColor}{RGB}{0,0,0}

\node[text=drawColor,anchor=base west,inner sep=0pt, outer sep=0pt, scale=  0.88] at ( 91.23,256.89) {Approximate Test};
\end{scope}
\begin{scope}
\path[clip] (  0.00,  0.00) rectangle (433.62,289.08);
\definecolor{drawColor}{RGB}{0,0,0}

\node[text=drawColor,anchor=base west,inner sep=0pt, outer sep=0pt, scale=  0.88] at ( 91.23,242.44) {Empirical Test};
\end{scope}
\end{tikzpicture}
	}
        \vspace{-6mm}
	\caption{Computation times (seconds). We can see that the
          Approximate Test is better than two orders of magnitude
          faster than the Empirical Test, but that they have the same
          asymptotic performance in both cases $O(|V|^2)$ }\label{fig:comp_1}
\end{figure}

\begin{table}[htbp]
	\centering
	\begin{tabular}{l|rrr}
                \multicolumn{4}{c}{Approximation Test} \\
		\hline
		& \multicolumn{3}{c}{$|V|$}\\
$\bar{d}$ & 100 & 1000 & 10000 \\ 
  \hline
  1 & 0.25 & 0.08 & 0.06 \\ 
    3 & 0.23 & 0.06 & 0.08 \\ 
    5 & 0.23 & 0.07 & 0.05 \\ 
   10 & 0.20 & 0.06 & 0.06 \\ 
   \hline

	\end{tabular}
        \hspace{20mm}
	\begin{tabular}{l|rrr}
                \multicolumn{4}{c}{Empirical Test} \\
		\hline
		& \multicolumn{3}{c}{$|V|$}\\
$\bar{d}$ & 100 & 316 & 1000 \\ 
  \hline
  1 & 0.05 & 0.03 & 0.07 \\ 
    3 & 0.06 & 0.04 & 0.07 \\ 
    5 & 0.07 & 0.07 & 0.04 \\ 
   10 & 0.06 & 0.07 & 0.03 \\ 
   \hline

	\end{tabular}
	\caption{Estimated significance levels obtained for a
          homogenous network.}
	\label{tab:sig_pv}
\end{table}


\subsection{Power}
\label{sub:Power}

Next we consider the \emph{power} of the methods, by simulating
networks that have controlled heterogeneity in the $p_{ij}$. We
considered a modification of the \erdos network which we call a
\hetero network (a modification of the models described in
\cite{Soederberg2003}). We modify the network to have two type of
nodes: denoted as red nodes and blue nodes. There will be $|V_1|$ red
nodes and $|V_2|$ blue nodes with $|V_1|$ not necessarily equal to
$|V_2|$. The total number of nodes in the network is $|V| = |V_1| +
|V_2|$. We also define, for $i > j$

\[
p_{ij} = P\{A_{i,j} = 1\} = \begin{cases}
p, &i,j\in V_1,\\
q, &i,j\in V_2,\\
\sqrt{pq}, & i \in V_1, j \in V_2,\\
\sqrt{pq}, & i \in V_2, j \in V_1.\\
\end{cases}
\]

The network is undirected so $A_{ij} = A_{ji}$ for $i < j$, and as always $A_{ii}=0$.

The choice of $\sqrt{pq}$ for the edges connecting nodes of different
classes was based on the idea that each node contributed to an edge
independently. To be consistent with $p_{ij} = p$ for $i, j\in V_1$, a
red node should contribute $\sqrt{p}$, and similarly a blue node
$\sqrt{q}$, and hence $p_{ij} = \sqrt{pq}$ for $i \in V_1, j \in V_2$.

We examine power using \hetero networks with larger networks (where
the Approximation Test is valid) and average node degrees as
considered before. In our case, we set $|V_1| = |V_2| = |V|/2$, \ie equal group sizes.

We control the degree of heterogenity through the ratio
\[ r = \frac{q - p}{p+q},
\]
which is the relative average difference between the two classes'
probabilities.  We test $r = \{0.01, 0.1, 0.2, 0.5, 0.75, 1\}$.  The
ratio $r$ is the ratio of the distance of $p$ and $q$ from the average
of $p$ and $q$ to the average of $p$ and $q$. The case $r=0$
corresponds to homogeneity, and values of $r$ greater than one would
result in either $p$ or $q$ being negative and hence are not possible.

For each parameter combination $(|V|,\bar{d}, r)$, we simulated 500
networks. For each we then calculated the P-value using
Algorithm~\ref{algo:gof_large}.  We have only considered the power for
the Approximate Test due to the excessive computation time of the
Empirical Test.

\begin{figure}[htbp]
	\centering
	\input{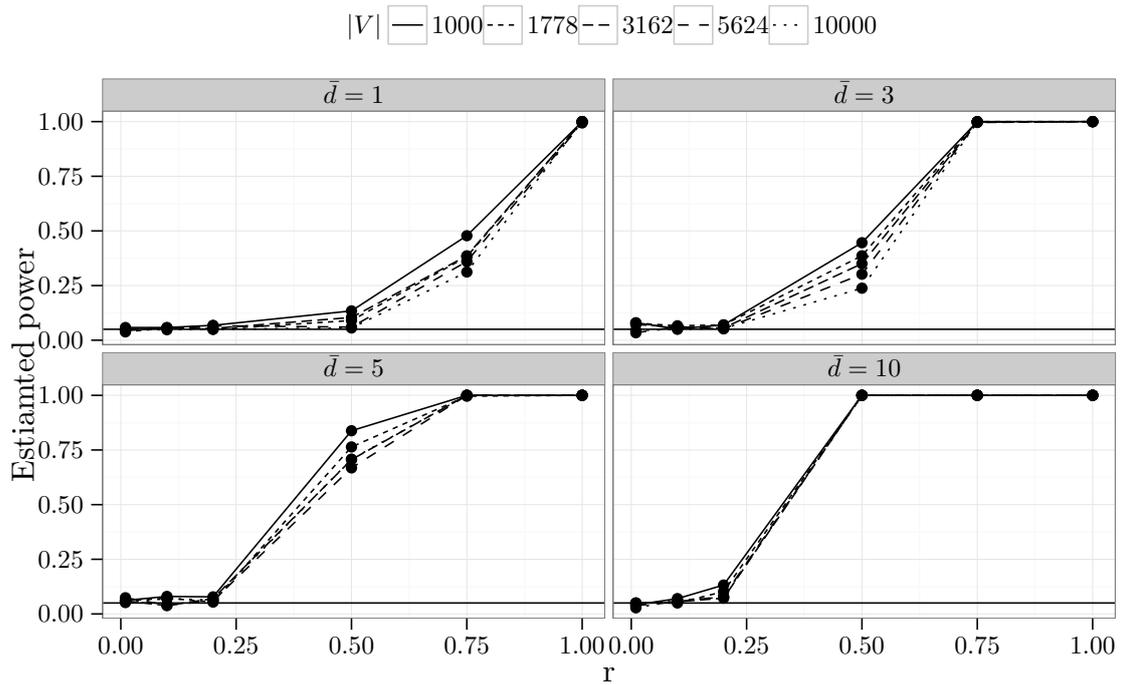}
        \vspace{-6mm}
	\caption{Estimated power for given \hetero networks with ratio
          $r$ faceted by average node degree for given network size
          $|V|$. Note that even for moderate $r$ values, the
          difference in probabilities being detected can be {\em very
            small}, \eg for $\bar{d} = 5$, $|V| = 1000$ and $r=0.5$,
          then $p - q = 0.005364$.}
	\label{fig:hetero_power}
\end{figure}

The results are given in Figure~\ref{fig:hetero_power}. As the ratio,
$r$, increases we observe a monotonically increasing power, \ie, the
probability of concluding that the \hetero network is inhomogeneous
increases. It converges to 1, as expected, because as the ratio
increases the difference between $p$ and $q$ for a given average value
of $p$ and $q$ increases and so we have more heterogenous
probabilities for the edges and so should be more likely to conclude
that the network is heterogenous.

The power of the test might not appear large for moderately small
values of $r$, but note that given the data, the test is actually very
sensitive. For example consider the case where $\bar{d} = 5$, $|V| =
1000$ and $r=0.5$. In this case, we will pick up 84\% of networks with
$p - q = 0.005364$: a very small difference in probabilities. We chose
to display the results using $r$ as the $x$-axis in order to make
clear comparisons, but that choice hides the difficulty of the
probabilities being tested.

As the average node degree $\bar{d}$ increases the estimated power
goes to 1 more quickly. This makes sense; networks with larger
$\bar{d}$ have on average more edges hence more information. For even
moderately non-sparse networks, the test is quite capable.

On the other hand, for a given ratio and $\bar{d}$, we observe that as
$|V|$ increases, the power decreases. This is because larger networks
with a given fixed average degree have smaller $p$ and $q$ values, and
hence a smaller difference in probabilities (for a given value of
$r$). The smaller difference is slightly harder to detect, even though
there are more data points, but note that by the time  average degree
$\bar{d} \approx 10$ the effect of network size is almost
negligable.

\subsection{Computational time}
\label{sec:Computational time}

We examine the computation times for the two methods on the test
networks described in Section~\ref{sub:Significance}. We show here the
results for networks with average node degree $\bar{d} = 5$.  We timed
each method 100 times on an iMac 2.7GHz Intel Core i5 with 4 cores and
16 GB of RAM.

The mean computation times for each method and network size is given
in Table~\ref{tab:comp_1} and Figure~\ref{fig:comp_1} from which we
see Empirical Test is orders of magnitude slower than Approximate
Test, but that both methods are $O(|V|^2)$.

\begin{table}[htbp]
	\centering
	\begin{tabular}{l|rrr}
		\hline
		& \multicolumn{3}{c}{$|V|$}\\
Method & 100 & 1000 & 10000 \\ 
  \hline
Approximate Test & 0.04s & 0.05s & 0.27s \\ 
  Empirical Test & 7.39s & 10.92s & 58.23s \\ 
   \hline

	\end{tabular}
	\caption{Average computation times (seconds) for the two methods for
          calculating P-values for networks with $|V|$ nodes, and $\bar{d}=5$. }
	\label{tab:comp_1}
\end{table}

\begin{figure}[htbp]
	\centering
\begin{tikzpicture}[x=1pt,y=1pt]
\definecolor{fillColor}{RGB}{255,255,255}
\path[use as bounding box,fill=fillColor,fill opacity=0.00] (0,0) rectangle (433.62,289.08);
\begin{scope}
\path[clip] (  0.00,  0.00) rectangle (433.62,289.08);
\definecolor{drawColor}{RGB}{255,255,255}
\definecolor{fillColor}{RGB}{255,255,255}

\path[draw=drawColor,line width= 0.6pt,line join=round,line cap=round,fill=fillColor] (  0.00,  0.00) rectangle (433.62,289.08);
\end{scope}
\begin{scope}
\path[clip] ( 40.27, 30.69) rectangle (428.12,283.58);
\definecolor{fillColor}{RGB}{255,255,255}

\path[fill=fillColor] ( 40.27, 30.69) rectangle (428.12,283.58);
\definecolor{drawColor}{gray}{0.98}

\path[draw=drawColor,line width= 0.6pt,line join=round] ( 40.27,122.60) --
	(428.12,122.60);

\path[draw=drawColor,line width= 0.6pt,line join=round] ( 40.27,215.14) --
	(428.12,215.14);

\path[draw=drawColor,line width= 0.6pt,line join=round] ( 67.69, 30.69) --
	( 67.69,283.58);

\path[draw=drawColor,line width= 0.6pt,line join=round] (165.63, 30.69) --
	(165.63,283.58);

\path[draw=drawColor,line width= 0.6pt,line join=round] (263.58, 30.69) --
	(263.58,283.58);

\path[draw=drawColor,line width= 0.6pt,line join=round] (361.52, 30.69) --
	(361.52,283.58);
\definecolor{drawColor}{gray}{0.90}

\path[draw=drawColor,line width= 0.2pt,line join=round] ( 40.27, 76.33) --
	(428.12, 76.33);

\path[draw=drawColor,line width= 0.2pt,line join=round] ( 40.27,168.87) --
	(428.12,168.87);

\path[draw=drawColor,line width= 0.2pt,line join=round] ( 40.27,261.41) --
	(428.12,261.41);

\path[draw=drawColor,line width= 0.2pt,line join=round] (116.66, 30.69) --
	(116.66,283.58);

\path[draw=drawColor,line width= 0.2pt,line join=round] (214.60, 30.69) --
	(214.60,283.58);

\path[draw=drawColor,line width= 0.2pt,line join=round] (312.55, 30.69) --
	(312.55,283.58);

\path[draw=drawColor,line width= 0.2pt,line join=round] (410.49, 30.69) --
	(410.49,283.58);
\definecolor{drawColor}{RGB}{0,0,0}
\definecolor{fillColor}{RGB}{0,0,0}

\path[draw=drawColor,line width= 0.4pt,line join=round,line cap=round,fill=fillColor] ( 57.90, 42.18) circle (  1.96);

\path[draw=drawColor,line width= 0.4pt,line join=round,line cap=round,fill=fillColor] (136.25, 46.12) circle (  1.96);

\path[draw=drawColor,line width= 0.4pt,line join=round,line cap=round,fill=fillColor] (214.60, 49.35) circle (  1.96);

\path[draw=drawColor,line width= 0.4pt,line join=round,line cap=round,fill=fillColor] (410.49, 57.91) circle (  1.96);

\path[draw=drawColor,line width= 0.4pt,line join=round,line cap=round,fill=fillColor] ( 57.90,256.19) circle (  1.96);

\path[draw=drawColor,line width= 0.4pt,line join=round,line cap=round,fill=fillColor] (136.25,260.33) circle (  1.96);

\path[draw=drawColor,line width= 0.4pt,line join=round,line cap=round,fill=fillColor] (214.60,263.72) circle (  1.96);

\path[draw=drawColor,line width= 0.4pt,line join=round,line cap=round,fill=fillColor] (410.49,272.08) circle (  1.96);

\path[draw=drawColor,line width= 0.6pt,line join=round] ( 57.90, 42.18) --
	(136.25, 46.12) --
	(214.60, 49.35) --
	(410.49, 57.91);

\path[draw=drawColor,line width= 0.6pt,dash pattern=on 2pt off 2pt ,line join=round] ( 57.90,256.19) --
	(136.25,260.33) --
	(214.60,263.72) --
	(410.49,272.08);
\definecolor{drawColor}{gray}{0.50}

\path[draw=drawColor,line width= 0.6pt,line join=round,line cap=round] ( 40.27, 30.69) rectangle (428.12,283.58);
\end{scope}
\begin{scope}
\path[clip] (  0.00,  0.00) rectangle (433.62,289.08);
\definecolor{drawColor}{RGB}{0,0,0}

\node[text=drawColor,anchor=base east,inner sep=0pt, outer sep=0pt, scale=  0.88] at ( 35.32, 73.30) {0.1};

\node[text=drawColor,anchor=base east,inner sep=0pt, outer sep=0pt, scale=  0.88] at ( 35.32,165.84) {1.0};

\node[text=drawColor,anchor=base east,inner sep=0pt, outer sep=0pt, scale=  0.88] at ( 35.32,258.38) {10.0};
\end{scope}
\begin{scope}
\path[clip] (  0.00,  0.00) rectangle (433.62,289.08);
\definecolor{drawColor}{RGB}{0,0,0}

\path[draw=drawColor,line width= 0.6pt,line join=round] ( 37.52, 76.33) --
	( 40.27, 76.33);

\path[draw=drawColor,line width= 0.6pt,line join=round] ( 37.52,168.87) --
	( 40.27,168.87);

\path[draw=drawColor,line width= 0.6pt,line join=round] ( 37.52,261.41) --
	( 40.27,261.41);
\end{scope}
\begin{scope}
\path[clip] (  0.00,  0.00) rectangle (433.62,289.08);
\definecolor{drawColor}{RGB}{0,0,0}

\path[draw=drawColor,line width= 0.6pt,line join=round] (116.66, 27.94) --
	(116.66, 30.69);

\path[draw=drawColor,line width= 0.6pt,line join=round] (214.60, 27.94) --
	(214.60, 30.69);

\path[draw=drawColor,line width= 0.6pt,line join=round] (312.55, 27.94) --
	(312.55, 30.69);

\path[draw=drawColor,line width= 0.6pt,line join=round] (410.49, 27.94) --
	(410.49, 30.69);
\end{scope}
\begin{scope}
\path[clip] (  0.00,  0.00) rectangle (433.62,289.08);
\definecolor{drawColor}{RGB}{0,0,0}

\node[text=drawColor,anchor=base,inner sep=0pt, outer sep=0pt, scale=  0.88] at (116.66, 19.68) {2.5};

\node[text=drawColor,anchor=base,inner sep=0pt, outer sep=0pt, scale=  0.88] at (214.60, 19.68) {5.0};

\node[text=drawColor,anchor=base,inner sep=0pt, outer sep=0pt, scale=  0.88] at (312.55, 19.68) {7.5};

\node[text=drawColor,anchor=base,inner sep=0pt, outer sep=0pt, scale=  0.88] at (410.49, 19.68) {10.0};
\end{scope}
\begin{scope}
\path[clip] (  0.00,  0.00) rectangle (433.62,289.08);
\definecolor{drawColor}{RGB}{0,0,0}

\node[text=drawColor,anchor=base,inner sep=0pt, outer sep=0pt, scale=  1.10] at (234.19,  7.70) {$\bar{d}$};
\end{scope}
\begin{scope}
\path[clip] (  0.00,  0.00) rectangle (433.62,289.08);
\definecolor{drawColor}{RGB}{0,0,0}

\node[text=drawColor,rotate= 90.00,anchor=base,inner sep=0pt, outer sep=0pt, scale=  1.10] at ( 15.28,157.13) {Computation time (sec)};
\end{scope}
\begin{scope}
\path[clip] (  0.00,  0.00) rectangle (433.62,289.08);
\definecolor{fillColor}{RGB}{255,255,255}

\path[fill=fillColor] (303.41,132.82) rectangle (397.69,181.45);
\end{scope}
\begin{scope}
\path[clip] (  0.00,  0.00) rectangle (433.62,289.08);
\definecolor{drawColor}{RGB}{0,0,0}

\node[text=drawColor,anchor=base west,inner sep=0pt, outer sep=0pt, scale=  1.10] at (307.68,169.61) {Method};
\end{scope}
\begin{scope}
\path[clip] (  0.00,  0.00) rectangle (433.62,289.08);
\definecolor{drawColor}{gray}{0.80}
\definecolor{fillColor}{RGB}{255,255,255}

\path[draw=drawColor,line width= 0.6pt,line join=round,line cap=round,fill=fillColor] (307.68,151.54) rectangle (322.13,165.99);
\end{scope}
\begin{scope}
\path[clip] (  0.00,  0.00) rectangle (433.62,289.08);
\definecolor{drawColor}{RGB}{0,0,0}

\path[draw=drawColor,line width= 0.6pt,line join=round] (309.12,158.77) -- (320.69,158.77);
\end{scope}
\begin{scope}
\path[clip] (  0.00,  0.00) rectangle (433.62,289.08);
\definecolor{drawColor}{gray}{0.80}
\definecolor{fillColor}{RGB}{255,255,255}

\path[draw=drawColor,line width= 0.6pt,line join=round,line cap=round,fill=fillColor] (307.68,137.08) rectangle (322.13,151.54);
\end{scope}
\begin{scope}
\path[clip] (  0.00,  0.00) rectangle (433.62,289.08);
\definecolor{drawColor}{RGB}{0,0,0}

\path[draw=drawColor,line width= 0.6pt,dash pattern=on 2pt off 2pt ,line join=round] (309.12,144.31) -- (320.69,144.31);
\end{scope}
\begin{scope}
\path[clip] (  0.00,  0.00) rectangle (433.62,289.08);
\definecolor{drawColor}{RGB}{0,0,0}

\node[text=drawColor,anchor=base west,inner sep=0pt, outer sep=0pt, scale=  0.88] at (323.94,155.74) {Approximate Test};
\end{scope}
\begin{scope}
\path[clip] (  0.00,  0.00) rectangle (433.62,289.08);
\definecolor{drawColor}{RGB}{0,0,0}

\node[text=drawColor,anchor=base west,inner sep=0pt, outer sep=0pt, scale=  0.88] at (323.94,141.28) {Empirical Test};
\end{scope}
\end{tikzpicture}
	\caption{Computation times (seconds) for each method for
          network of given average node degree $\bar{d}$ and
          $|V|=1000$. We can observe that the times are $O(|E|)$. }
	\label{fig:comp_2}
\end{figure}
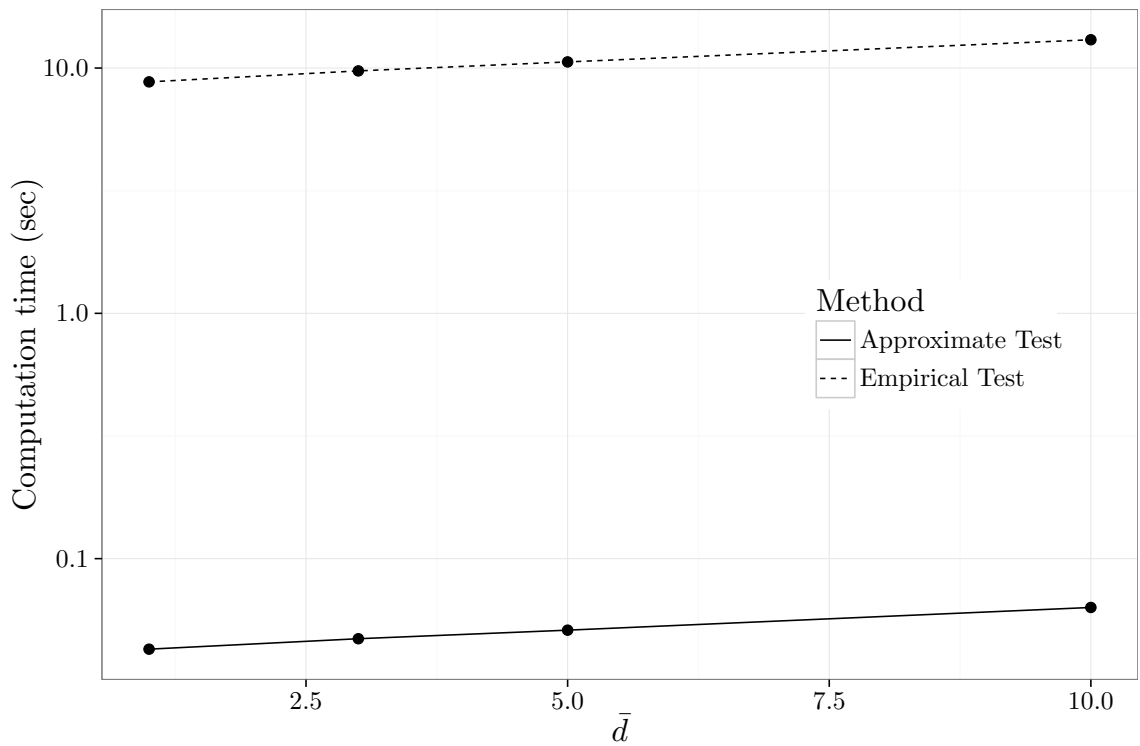

We examine the effect of average node degree on the computation time
in Table~\ref{tab:comp_2} and Figure~\ref{fig:comp_2}. Again, we can
see that as expected the Empirical Test is more than two orders of
magnitude slower than the Approximate Test. The computational cost
increases with $\bar{d}$ as $O(|E|)$.

\begin{table}[htbp]
	\centering
	\begin{tabular}{l|rrrr}
		\hline
		& & \\[-10pt]
		& \multicolumn{4}{c}{$\bar{d}$}\\
Method & 1 & 3 & 5 & 10 \\ 
  \hline
Approximate Test & 0.04 & 0.05 & 0.05 & 0.06 \\ 
  Empirical Test & 8.78 & 9.74 & 10.59 & 13.04 \\ 
   \hline

	\end{tabular}
	\caption{Average computation times (seconds) for the two
          methods for networks with average node degree $\bar{d}$ and
          $|V|=1000$.}
	\label{tab:comp_2}
\end{table}

\clearpage
\section{Application to Australian Research Council data}

We will now apply the homogeneity goodness-of-fit test to a real
dataset obtained from the Australian Research
Council\footnote{\url{http://www.arc.gov.au/}}. This data consists of
the Field of Research (FoR) codes entered by applications for all
Discovery Project grants applied for between 2010 and 2014
inclusively.

The FoR codes are six-digit number that indicates the areas of
research covered by the grant, as outlined by the Australian Bureau of
Statistics (ABS)\footnote{\url{http://goo.gl/hrWMUh}}. The codes are
hierarchical: that is they consists of three nested 2-digit codes, for
example, 010406 is Stochastic Analysis and Modelling which is nested
within 0104 (Statistics) which is contained within 01 (Mathematical
Sciences). There are 1238 possible 6-digit FoR codes, though only 1174
were used in the dataset.

Each of the 18,476 grant applications in the dataset nominated one or
more FoR codes. These are used to help selection of reviewers, and for
statistical purposes (for instance, to report the number of grants
accepted per research area).

\begin{figure}[htbp]
	\centering
	\includegraphics[width=0.45\textwidth]{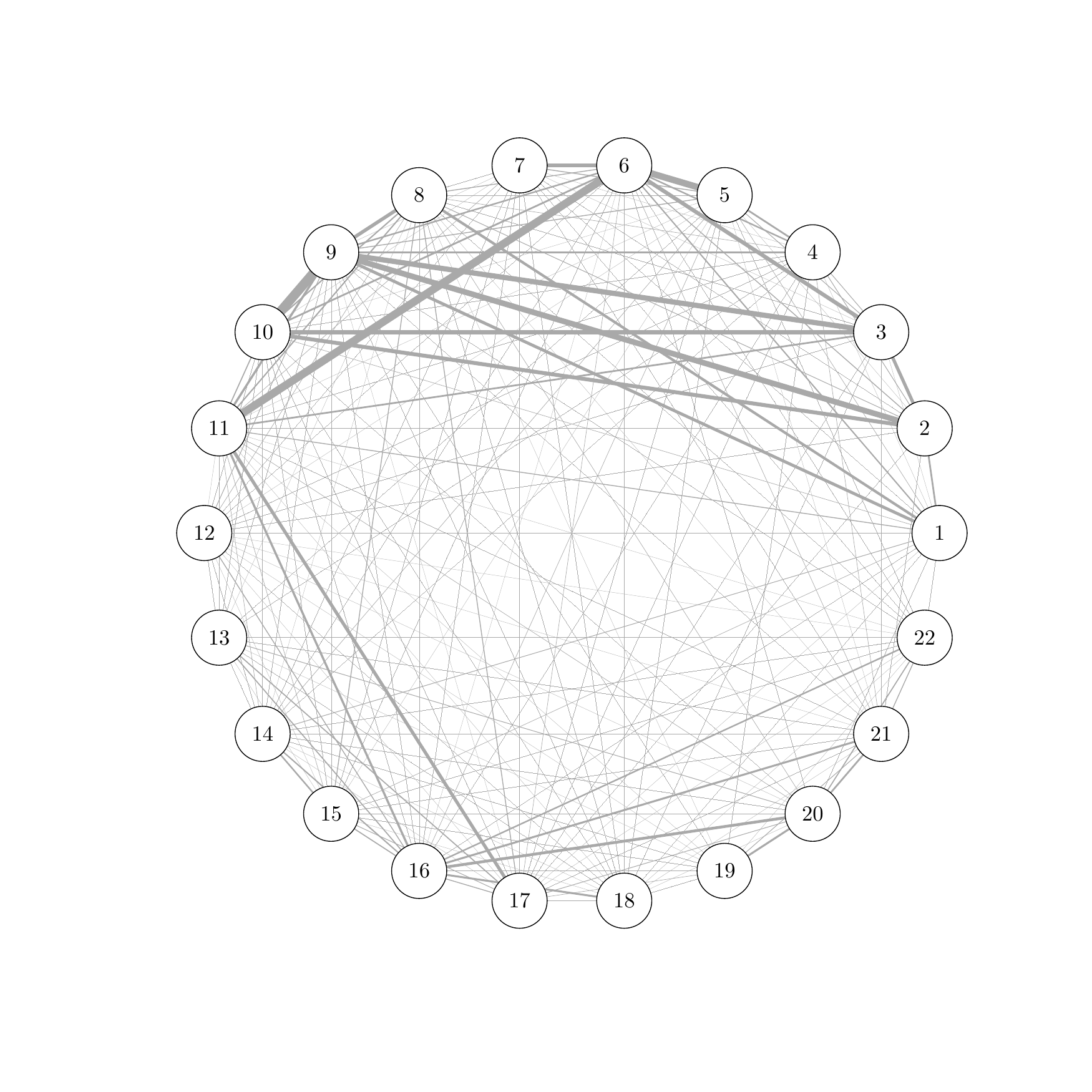}
        \vspace{-6mm}
	\caption{ARC 2-digit FoR-code graph: shows
          cross-collaborations between 2-digit FoR groupings. Note
          that at this level the graph is almost complete. We actually
        analyse the 6-digit graph which is much sparser, but it has
        too many nodes to usefully visualise. We use the width of the
        lines to illustrate the number of common grant applications.}
	\label{fig:arc_graph}
\end{figure}

Many grant applications nominate more than one FoR code (the largest
number in a single grant was 11).  We construct an inter-FoR-code
network by creating a link between two FoR codes if there was at least
one application that contained both of these FoR
codes. Figure~\ref{fig:arc_graph} shows the 2-digit graph as an
illustration of the data, though we analyse the 6-digit graph
here. This gave us a network with 1238 nodes and 15,747 edges.

We have enough nodes (\ie $|V| = 1,238 \geq 1000$) to use the
Approximation Test to assess homogeneity. We sampled 1000 subgraphs of
size 876 nodes. This gave an observed $X^2$ test statistic of
$21,226.12$ with a P-value of 0. Therefore, we conclude that we have
very strong evidence that there is heterogeneity in the 6-digit
FoR-code network.

This is not at all surprising: intuition suggests that
cross-collaboration between different fields is dependent on the type
of field under consideration. Certain fields are often used within
others, for instance Statistics is an important component of many
other areas of Science, and hence there is likely to be many bridges
between Statistics and Sciences. On the other hand, subjects such as
Pure Mathematics exist in relative isolation as a deliberate
decision about the nature of the subject. However, while the finding
is not surprising, it is reassuring to have a quantitative test,
rather than relying on intuition.


\section{Discussion}

In the beginning, we discussed a problem in network analysis, \ie,
that we often have a single observed network (which is a single
high-dimensional data-point) from which we would like to choose an
appropriate model, and then estimate the parameters for the
model. This is often achieved by implicitly assuming homogeneity of
the subgraph structure of the network so that subsamples can be
exploited to provide multiple data points in model selection or
parameter estimation.

We have introduced a simple example of proof of concept of a framework
to access this assumption. However our framework can be generalised to
elucidate other notions of heterogeneity in the subgraphs.

Our example of using this framework was the concept of
homogeneity/heterogeneity in edge probability. We used the number of
edges as a simple summary statistic of subsampled graphs. In this
example, we have obtained excellent results.

Testing that a network in \erdos is not that remarkable. It is more the
philosophy of the approach -- we have proposed a framework to separate local
structure from global structure. That is, we are not aiming to test if
the network is a  \erdos random graph, but rather we aim to test a
property of the network that would be needed if we were to fit the
\erdos random graph to the data.

So how can this framework be applied to other forms of heterogeneity?
The outline of the approach is as follows:
\begin{itemize}
\item Identify the property that is being explored (in our case edge
  connection probability), and identify an appropriate model with this
  property: in our case the GER random graphs.

\item Identify summary statistics for the model parameter. Again in
  our case, we choose the number of edges in the subgraph to give the
  most information about the parameter $p_{ij}$.

\item Sample subgraphs from the network and for each one record the
  summary statistic.

\item Perform a goodness of fit test for the summary statistic
  compared to the theoretical distribution of the summary
  statistic. The theoretical distribution can be derived from first
  principles. Or instead, you can derive a empirical distribution for
  the summary statistic under the the assumption of homogeneity by
  simulating the appropriate network and the sampling from this.

\end{itemize}
Given this framework, we can now test for heterogeneity of the local
network structure, and access whether a single homogenous model can be
applied to the total network, though choosing appropriate sampling and
summary statistics might require some creativity for more complex
inferences.


\section{Conclusion}
\label{sec:Conclusion}
In this paper, we have illustrated how a sampling procedure can be used to test for heterogeneity in the local structure of a network. We used the simple example of heterogeneity in edge probability of an \erdos network as a proof of concept, but the method can be applied to other heterogeneities.

We have shown how a goodness of fit can be utilised to test for heterogenity. In our example we use the number of edges in the subgraphs as the summary statistic and for this we could calculate an approximate distribution using the hypergeometric. An appropriate distribution may not always be available and hence we also shown how an empirical distribution can also be used.

This framework gives the researcher a new method to ellucidate heterogeneity in local structure.

An area of future research that we are investigating is using this approach to
not only indicate heterogeneity, but to identify its relationship to network
structure: a residual type object for network modelling.

\section*{Acknowledgements}

This work was partially supported by the Australian Research Council
through grant DP110103505, and the ARC Centre of Excellence for
Mathematical \& Statistical Frontiers.

We would also like to thank Aidan Byrne and the ARC for the provision
of the For-code dataset used here.

\bibliographystyle{plainnat}
\bibliography{homo_graph}

\appendix

\clearpage
\section{Algorithm to tabulate observed number of edges}
\label{sec:Algorithm to tabulate observed number of edges}

\begin{algorithm}[htbp]
	\LinesNumbered
	\SetKwInOut{Input}{input}
	\SetKwInOut{Output}{output}
	\Input{Parameters for hypergeometric distribution to test for:\\
	$\quad m$ number of successes in population.\\
	$\quad n$ number of failures in population.\\
	$\quad k$ sample size.\\
	$N$ Number of observations}
	\BlankLine

	$c \leftarrow 5/N$

	$p \leftarrow c$

	$j \leftarrow 1$

	\While{$p + c < 1 - c$}{
	$x_j \leftarrow P(X \leq x_j) = p$ where $X \sim hyper(m,n,k)$

	$p \leftarrow P(X \leq x_j) + c$

	$ j \leftarrow j + 1$
	}
	\Return{$\{x_1,\ldots\}$}
	\caption{Method to get bins with expected value of 5 or greater according to $hyper(m,n,k)$}
	\label{algo:bin}
\end{algorithm}

\end{document}